\documentclass[aps,prd,onecolumn,groupedaddress,showpacs,nofootinbib,amssymb]{revtex4}
\usepackage{graphicx,bm,color}
\usepackage{amsmath}
\usepackage{amssymb}
\usepackage{amsfonts}

\newcommand{\be}{\begin{equation}}
\newcommand{\ee}{\end{equation}}
\newcommand{\bea}{\begin{eqnarray}}
\newcommand{\eea}{\end{eqnarray}}
\newcommand{\beaa}{\begin{eqnarray*}}
\newcommand{\eeaa}{\end{eqnarray*}}

\newcommand{\nn}{\nonumber \\}
\newcommand{\e}{\mathrm{e}}

\allowdisplaybreaks[4]

\begin{document}

\title{Mimetic $F(R)$ gravity: inflation, dark energy and bounce}

\author{Shin'ichi Nojiri$^{1,2}$ and Sergei D. Odintsov$^{3}$\footnote{
Also, at King Abdulaziz University, Jeddah, Saudi Arabia
}}

\affiliation{ $^1$ Department of Physics, Nagoya University, Nagoya
464-8602, Japan \\
$^2$ Kobayashi-Maskawa Institute for the Origin of Particles and
the Universe, Nagoya University, Nagoya 464-8602, Japan \\
$^3$Instituci\`{o} Catalana de Recerca i Estudis Avan\c{c}ats
(ICREA) and Institut de Ciencies de l'Espai (IEEC-CSIC), C5, Campus UAB, 08193
Bellaterra (Barcelona), Spain}

\begin{abstract}

We propose the   mimetic $F(R)$ theory and investigate the early-time and 
late-time acceleration in such theory.
It is demonstrated that inflation consistent with observable data may be 
realized in such
theory.
The reconstruction of realistic $\Lambda$CDM era is also possible as well as 
the
unification of early-time inflation with late-time acceleration or bounce 
universe. It is stressed that specific universe evolution is governed by 
mimetic $F(R)$ theory which is different from convenient $F(R)$ gravity. The 
corresponding examples are presented.
Mimetic $F(R)$ gravity is generalized by the addition of the scalar potential 
in the
formulation of convenient $F(R)$ gravity with specific Lagrange multiplier 
constraint. It is
demonstrated that such theory may admit the arbitrary universe evolution via 
the
corresponding choice of the scalar potential and/or function $F(R)$.

\end{abstract}

\pacs{11.30.-j, 95.36.+x, 98.80.Cq}

\maketitle

\section{Introduction}

During last years it has been proven that modified gravity (for a recent
review, see \cite{review}) is quite successful in description of
the early-time and late-time acceleration of the universe. Number of viable 
modified
gravity theories has been elaborated.
One of the most interesting models of that sort is $F(R)$ gravity which is 
known to be
ghost-free theory like much more complicated ghost-free massive gravity (for 
recent
review, see \cite{review1}).

Recently, new approach to GR has been developed in such a way that it respects
conformal symmetry as internal degree of freedom.
Usually we regard the metric $g_{\mu\nu}$ as a fundamental variable of the
gravity.
It is, however, possible to express the metric in a different way by using new
degrees of freedom.
In such an expression, the equation given by the new degrees of freedom may
admit a new or wider class of solutions if compare with the equations given by 
the
variation over the metric.
An example of such  models is the mimetic model \cite{Chamseddine:2013kea}
(for its generalizations, see \cite{Chamseddine:2014vna,mimetic}).
In the mimetic model, we parametrize the metric in the following form.
\be
\label{Mi1}
g_{\mu\nu}= - \hat g^{\rho\sigma} \partial_\rho \phi \partial_\sigma \phi
\hat g_{\mu\nu} \, .
\ee
Instead of considering the variation of the action with respect to
$g_{\mu\nu}$, we consider the variation with respect to $\hat g_{\mu\nu}$ and 
$\phi$.
Because the parametrization is invariant under the Weyl transformation
$\hat g_{\mu\nu} \to \e^{\sigma(x)} \hat g_{\mu\nu}$, the variation over
$\hat g_{\mu\nu}$ gives the traceless part of the equation.
In fact, in case of the Einstein gravity, whose action is given by
\be
\label{EHaction}
S=\int d^4 x \sqrt{-g\left({\hat g}_{\mu\nu}, \phi \right)}
\left( \frac{R\left({\hat g}_{\mu\nu}, \phi \right)}{2\kappa^2} +
\mathcal{L}_\mathrm{matter}\right)\, ,
\ee
the variation over $\hat g_{\mu\nu}$ gives the traceless part of the Einstein 
equation:
\be
\label{Mi2}
0 = - R\left({\hat g}_{\mu\nu}, \phi \right)_{\mu\nu}
+ \frac{1}{2} g\left({\hat g}_{\mu\nu}, \phi \right)_{\mu\nu}
R\left({\hat g}_{\mu\nu}, \phi \right) + \kappa^2 T_{\mu\nu}
+ \partial_\mu \phi \partial_\nu \phi \left( R\left({\hat g}_{\mu\nu}, \phi
\right) + \kappa^2 T \right)\, .
\ee
Here $T$ is the trace of the matter energy-momentum tensor $T_{\mu\nu}$,
$T=g\left({\hat g}_{\mu\nu}, \phi \right)^{\mu\nu}T_{\mu\nu}$.
Eq.~(\ref{Mi1}) shows
\be
\label{Mi2B}
g\left({\hat g}_{\mu\nu}, \phi \right)^{\mu\nu} \partial_\mu \phi
\partial_\nu \phi = - 1\, .
\ee
freedom,
Usually, the variation over the Weyl factor in the metric gives the equation
for the trace part, that is, in case of the Einstein gravity,
$R + \kappa^2 T =0$ but due to the parametrization in (\ref{Mi1}), the 
variation over
$\phi$ gives
\be
\label{Mi3}
0 = \nabla\left(g\left({\hat g}_{\mu\nu}, \phi \right)_{\mu\nu}\right)^\mu
\left(\partial_\mu \phi \left( R\left({\hat g}_{\mu\nu}, \phi \right)
+ \kappa^2 T \right) \right) \, .
\ee
Note that $\nabla_\mu$ is the covariant derivative with respect to 
$g_{\mu\nu}$.
Eq.~(\ref{Mi3}) shows that there can be a wider class of the solutions in the
mimetic model if compare with the Einstein gravity.
In fact, Eq.~(\ref{Mi3}) effectively induces dark matter 
\cite{Chamseddine:2013kea}, which
will be discussed in the following section in case of the $F(R)$ extension of 
the mimetic
model.
We should note that in Eqs.~(\ref{Mi2}) and (\ref{Mi3}), ${\hat g}_{\mu\nu}$
appears only in the combination of $g_{\mu\nu}$ in (\ref{Mi1}) and therefore
${\hat g}_{\mu\nu}$ does not appear explicitly.

In the present letter we propose new model: mimetic $F(R)$ gravity.
This theory seems to be  ghost-free like the conventional $F(R)$ gravity and
conformally-invariant one.
We investigate the early-time and late-time accelerated universe in mimetic 
$F(R)$
gravity.
It is demonstrated that inflation consistent with observable data may be 
realized within
such approach. Furthermore, the reconstruction of LCDM model is also possible 
as well as
unification of early-time inflation with late-time acceleration in the spirit 
of first proposal
in the original $F(R)$ gravity \cite{uni}.
The example of bounce universe is also constructed. It is important to note 
that same cosmological evolution is realized by different form of the function 
$F(R)$ in mimetic theory if compare with convenient $F(R)$ gravity. The 
explicit examples of such different forms of $F(R)$ are given.
Like usual mimetic gravity, the theory under consideration admits the 
generalization by
adding the scalar potential in the formulation with Lagrange multiplier.
The accelerating cosmology for such theory is briefly discussed.

\section{Accelerating universe from mimetic $F(R)$ gravity}

Let us start from $F(R)$ gravity, whose action is given by
\be
\label{FRaction}
S=\int d^4 x \sqrt{-g} \left( F\left(R\right)
+ \mathcal{L}_\mathrm{matter}\right)\, .
\ee
Here $F(R)$ is some function of the Ricci scalar $R$ and
$\mathcal{L}_\mathrm{matter}$ is matter Lagrangian.
If we parametrize the metric as in (\ref{Mi1}), the variation of the metric is
given by
\be
\label{dg}
\delta g_{\mu\nu} = \hat g^{\rho\tau} \delta \hat g_{\tau\omega}
\hat g^{\omega\sigma} \partial_\rho \phi \partial_\sigma \phi \hat g_{\mu\nu}
  - \hat g^{\rho\sigma} \partial_\rho \phi \partial_\sigma \phi \delta  \hat 
g_{\mu\nu}
  - 2 \hat g^{\rho\sigma} \partial_\rho \phi \partial_\sigma \delta \phi \hat 
g_{\mu\nu} \, .
\ee
Then in case of $F(R)$ gravity, by using the parametrization of the metric as
in (\ref{Mi1}),
\be
\label{miFRaction}
S=\int d^4 x \sqrt{-g\left({\hat g}_{\mu\nu}, \phi \right)}
\left( F\left(R\left({\hat g}_{\mu\nu}, \phi \right)\right)
+ \mathcal{L}_\mathrm{matter}\right)\, ,
\ee
the variation of the action is given by
\begin{align}
\label{deltaS}
\delta S =& \int d^4 x \sqrt{-g\left({\hat g}_{\mu\nu}, \phi \right)}
\left(  \hat g^{\rho\tau} \delta \hat g_{\tau\omega} \hat g^{\omega\sigma}
\partial_\rho \phi \partial_\sigma \phi \hat g^{\mu\nu}
  - \hat g^{\rho\sigma} \partial_\rho \phi \partial_\sigma \phi
\hat g^{\mu\tau} \delta \hat g_{\tau\omega} \hat g^{\omega\nu}
  - 2 \hat g^{\rho\sigma} \partial_\rho \phi \partial_\sigma \delta \phi
\hat g^{\mu\nu}
\right) \left( \hat g^{\rho\sigma} \partial_\rho \phi \partial_\sigma \phi
\right)^{-2} \nn
&\times \left( \frac{1}{2} g_{\mu\nu} F\left(R\left({\hat g}_{\mu\nu}, \phi
\right)\right) - R\left({\hat g}_{\mu\nu}, \phi \right)_{\mu\nu}
F'\left(R\left({\hat g}_{\mu\nu}, \phi \right)\right)
+ \nabla\left(g\left({\hat g}_{\mu\nu}, \phi \right)_{\mu\nu}\right)_\mu
\nabla\left(g\left({\hat g}_{\mu\nu}, \phi \right)_{\mu\nu} \right)_\nu
F'\left(R\left({\hat g}_{\mu\nu}, \phi \right)\right) \right. \nn
& \left. - g\left({\hat g}_{\mu\nu}, \phi \right)_{\mu\nu}
\Box \left({\hat g}_{\mu\nu}, \phi \right)
F'\left(R\left({\hat g}_{\mu\nu}, \phi \right)\right) + \frac{1}{2} T_{\mu\nu} 
\right) \nn
= & \int d^4 x \sqrt{-g\left({\hat g}_{\mu\nu}, \phi \right)}
\left\{   - g^{\mu\tau} \delta \hat g_{\tau\omega} g^{\omega\nu} \right. \nn
& \times \left( \partial_\mu \phi \partial_\nu \phi
\left( 2 F\left(R\left({\hat g}_{\mu\nu}, \phi \right)\right)
  - R\left({\hat g}_{\mu\nu}, \phi \right)
F'\left(R\left({\hat g}_{\mu\nu}, \phi \right)\right)
  - 3 \Box\left(g\left({\hat g}_{\mu\nu}, \phi \right)_{\mu\nu}\right)
F'\left(R\left({\hat g}_{\mu\nu}, \phi \right)\right)+ \frac{1}{2} T \right) 
\right. \nn
& + \frac{1}{2} g_{\mu\nu} F\left(R\left({\hat g}_{\mu\nu}, \phi \right)\right)
  - R\left({\hat g}_{\mu\nu}, \phi \right)_{\mu\nu}
F'\left(R\left({\hat g}_{\mu\nu}, \phi \right)\right)
+ \nabla\left(g\left({\hat g}_{\mu\nu}, \phi \right)_{\mu\nu}\right)_\mu
\nabla\left(g\left({\hat g}_{\mu\nu}, \phi \right)_{\mu\nu} \right)_\nu
F'\left(R\left({\hat g}_{\mu\nu}, \phi \right)\right) \nn
& \left. - g\left({\hat g}_{\mu\nu}, \phi \right)_{\mu\nu}
\Box \left({\hat g}_{\mu\nu}, \phi \right)
F'\left(R\left({\hat g}_{\mu\nu}, \phi \right)\right) + \frac{1}{2} T_{\mu\nu} 
\right) \nn
& + 2 \delta \phi \nabla\left(g\left({\hat g}_{\mu\nu}, \phi \right)_{\mu\nu} 
\right)^\mu
\left(\partial_\mu \phi \left( 2 F\left(R\left({\hat g}_{\mu\nu}, \phi 
\right)\right)
  - R\left({\hat g}_{\mu\nu}, \phi \right) F'\left(R\left({\hat g}_{\mu\nu},
\phi \right)\right) \right. \right. \nn
& \left. \left. \left. - 3 \Box\left(g\left({\hat g}_{\mu\nu}, \phi 
\right)_{\mu\nu}\right)
F'\left(R\left({\hat g}_{\mu\nu}, \phi \right)\right) + \frac{1}{2} T \right) 
\right) \right\}\, .
\end{align}
Hence, the $F(R)$ equation corresponding to (\ref{Mi2}) has the following form:
\begin{align}
\label{Mi4}
0 = & \frac{1}{2} g_{\mu\nu} F\left(R\left({\hat g}_{\mu\nu}, \phi 
\right)\right)
  - R\left({\hat g}_{\mu\nu}, \phi \right)_{\mu\nu}
F'\left(R\left({\hat g}_{\mu\nu}, \phi \right)\right)
+ \nabla\left(g\left({\hat g}_{\mu\nu}, \phi \right)_{\mu\nu}\right)_\mu
\nabla\left(g\left({\hat g}_{\mu\nu}, \phi \right)_{\mu\nu} \right)_\nu
F'\left(R\left({\hat g}_{\mu\nu}, \phi \right)\right) \nn
& - g\left({\hat g}_{\mu\nu}, \phi \right)_{\mu\nu}
\Box \left({\hat g}_{\mu\nu}, \phi \right)
F'\left(R\left({\hat g}_{\mu\nu}, \phi \right)\right) + \frac{1}{2} T_{\mu\nu} 
\nn
& + \partial_\mu \phi \partial_\nu \phi
\left( 2 F\left(R\left({\hat g}_{\mu\nu}, \phi \right)\right)
  - R\left({\hat g}_{\mu\nu}, \phi \right)
F'\left(R\left({\hat g}_{\mu\nu}, \phi \right)\right)
  - 3 \Box\left(g\left({\hat g}_{\mu\nu}, \phi \right)_{\mu\nu}\right)
F'\left(R\left({\hat g}_{\mu\nu}, \phi \right)\right)+ \frac{1}{2} T \right) \, 
.
\end{align}
On the other hand, the equation corresponding to (\ref{Mi3}) has the following 
form:
\be
\label{Mi5}
0 = \nabla\left(g\left({\hat g}_{\mu\nu}, \phi \right)_{\mu\nu}\right)^\mu
\left(\partial_\mu \phi \left( 2 F\left(R\left({\hat g}_{\mu\nu}, \phi 
\right)\right)
  - R\left({\hat g}_{\mu\nu}, \phi \right) F'\left(R\left({\hat g}_{\mu\nu},
\phi \right)\right) - 3 \Box\left(g\left({\hat g}_{\mu\nu}, \phi 
\right)_{\mu\nu}\right)
F'\left(R\left({\hat g}_{\mu\nu}, \phi \right)\right) + \frac{1}{2} T \right) 
\right)\, .
\ee
In Eqs.~(\ref{Mi4}) and (\ref{Mi5}), ${\hat g}_{\mu\nu}$ appears only in the
combination of $g_{\mu\nu}$ in (\ref{Mi1}) and therefore ${\hat g}_{\mu\nu}$ 
does not
appear explicitly.
Then in the following, we abbreviate the ${\hat g}_{\mu\nu}$ and $\phi$
dependence of $g_{\mu\nu}$ and $R$.
We should note that any solution of the standard $F(R)$ gravity is also a
solution of the mimetic $F(R)$ gravity.
This is because in the standard $F(R)$ gravity, Eq.~(\ref{Mi5}) is always
satisfied since we find $ 2 F(R) - R F'(R) - 3 \Box F'(R) + \frac{1}{2} T=0$ in 
the standard $F(R)$
gravity by considering the trace part of the equation in the $F(R)$ gravity.
The mimetic $F(R)$ gravity is ghost-free and conformally invariant theory.
We should note again that in (\ref{Mi4}) and (\ref{Mi5}), ${\hat g}_{\mu\nu}$
appears only in the combination of $g_{\mu\nu}$ in (\ref{Mi1}) and therefore
${\hat g}_{\mu\nu}$ does not appear explicitly.

We now assume the FRW space-time with flat spatial part,
\be
\label{FRW}
ds^2 = - dt^2 + a(t)^2 \sum_{i=1,2,3} {dx^i}^2 \, ,
\ee
where $R=6\dot H + 12 H^2$ and $\phi$ only depends on time $t$.
Due to Eq.~(\ref{Mi2B}), we find $\phi=t$.
Then Eq.~(\ref{Mi5}) gives
\begin{align}
\label{Mi5b}
\frac{C_\phi}{a^3} = & 2 F(R) - R F'(R) - 3 \Box F'(R) + \frac{1}{2} T \nn
= & 2 F(R) - 6 \left( \dot H + 2 H^2 \right) F'(R) + 3 \frac{d^2 F'(R)}{dt^2}
+ 9H \frac{d F'(R)}{dt} + \frac{1}{2} \left( - \rho + 3 p \right) \, .
\end{align}
Here $C_\phi$ is a constant.
Then in the second line of Eq.~(\ref{Mi4}), only $(t,t)$ component does not
vanish and behaves as $a^{-3}$ and therefore the solution of Eq.~(\ref{Mi5b}) 
with
$C_\phi\neq 0$ plays a role of the mimetic dark matter as in 
\cite{Chamseddine:2013kea}.

On the other hand the $(t,t)$ and $(i,j)$-components in (\ref{Mi4}) give the
identical equation:
\be
\label{Mi6}
0 = \frac{d^2 F'(R)}{dt^2} + 2H \frac{dF'(R)}{dt} - \left(\dot H + 3 H^2
\right) F'(R) + \frac{1}{2} F(R) + \frac{1}{2} p \, .
\ee
By combining (\ref{Mi5b}) and (\ref{Mi6}), we obtain
\be
\label{Mi7}
0 = \frac{d^2 F'(R)}{dt^2} - H \frac{d F'(R)}{dt} + 2 \dot H F'(R) + 
\frac{1}{2} \left(
p + \rho \right) + \frac{4C_\phi}{a^3}\, .
\ee
When $C_\phi=0$, the above equations reduce to those in the standard $F(R)$ 
gravity, or in
other words, when $C_\phi\neq 0$, the equation and therefore the solutions are 
different from those in the standard $F(R)$ gravity.
Note that when $\rho=p=C_\phi=0$, the de Sitter or anti-de Sitter
solution, where the scalar curvature $R$ is a constant $R=R_0 = 12 H_0^2$
$(H=H_0)$, is always a solution if the following equation given by (\ref{Mi6}) 
is satisfied,
\be
\label{dS1}
0 = - 6H_0^2 F'\left(12 H_0^2 \right) + F\left(12 H_0^2 \right) \, .
\ee
This situation is not changed from the standard $F(R)$ gravity.

Now one may study an arbitrary evolution of the scale factor $a(t)$ and
assume the explicit $a(t)$ dependence of $\rho$ and $p$ as for usual perfect 
fluid.
Let us consider the following differential equation:
\be
\label{Mi8}
0 = \frac{d^2 f(t)}{dt^2} - H(t) \frac{d f(t)}{dt} + 2 \dot H(t) f(t)\, .
\ee
Let the two solutions of (\ref{Mi8}) be $f_1(t)$ and $f_2(t)$.
Then $F'(R)$, which is the solution of (\ref{Mi7}), is given by
\begin{align}
\label{Mi9}
F'\left(R \left(t\right)\right) =& - f_1 (t) \int^t dt' \frac{ \gamma(t')
f_2(t') }{W\left(f_1(t'),f_2(t')\right)}
+ f_2 (t) \int^t dt' \frac{ \gamma(t') f_1(t') }{
W\left(f_1(t'),f_2(t')\right)} \, , \nn
\gamma(t) \equiv & - \frac{1}{2} \left( p + \rho \right) - \frac{4 
C_\phi}{a^3}\, , \nn
W\left(f_1,f_2\right) \equiv & f_1(t) f_2'(t) - f_2(t) f_1'(t)\, .
\end{align}
By using the $t$ dependence of the scalar curvature
$R=R(t)$, we find $t$ as a function of $R$, $t=t(R)$.
Hence, one gets the explicit form of $F'(R)$.

As an explicit example, we consider
\be
\label{Mi10}
H = \frac{H_0}{1 + \epsilon H_0 t}\, ,
\ee
which gives a constant slow roll parameter $\epsilon= - \frac{\dot H}{H^2}$.
Then
\be
\label{Mi11}
f_{1,2}(t) = f_\pm (t) = f_\pm^{(0)} \left( 1 + \epsilon H_0 t
\right)^{\alpha_\pm}\, ,\quad
\alpha_\pm \equiv \frac{1 + \frac{1}{\epsilon} \pm \sqrt{ \left( 1
+ \frac{1}{\epsilon}\right)^2 + \frac{8}{\epsilon}}}{2}\, .
\ee
Here $f_\pm^{(0)}$'s are constants.
We now assume $p$ and $\rho$ are given by the perfect fluid whose equation of
state parameter is $w$, that is, $p=w\rho$.
Then by using (\ref{Mi9}), we find
\be
\label{Mi12}
F(R) = A R^{ - \frac{3(1+w)}{2\epsilon}} + B R^{ - \frac{3}{2\epsilon}}
+ C_+ R^{- \frac{\alpha_+}{2}} + C_- R^{- \frac{\alpha_-}{2}} \, .
\ee
Here $A$, $B$, and $C_\pm$ are constant.
The constant $A$ is proportional to the energy density and $B$ is proportional 
to $C_\phi$
but $C_\pm$ can be arbitrary and as a special case $C_\pm$ may vanish.
Therefore in the standard $F(R)$ gravity $B=0$ but in the mimetic $F(R)$ 
theory, $B$ should
not vanish.
Then if we do not include the matter, that is, $A=0$, the simplest action in 
the standard $F(R)$
gravity could be
\be
\label{MiA1}
F(R) = C_+ R^{- \frac{\alpha_+}{2}} \ \mbox{or}\ F(R) =  C_- R^{- 
\frac{\alpha_-}{2}} \, .
\ee
but the simplest action in the mimetic $F(R)$ gravity could be
\be
\label{MiA2}
F(R) = B R^{ - \frac{3}{2\epsilon}} \, .
\ee
Therefore for the identical evolution of the universe, the corresponding forms 
of $F(R)$ can be
different in the standard $F(R)$ gravity and the mimetic $F(R)$ gravity, in 
general.

Other slow roll parameters are given by
\be
\label{Mi12B}
\eta \sim \frac{3\dot H}{2H^2} = \frac{3}{2}\epsilon\, , \quad
\xi^2 \sim - \frac{3 {\dot H}^2}{2H^4} = \frac{3}{2}\epsilon^2\, ,
\ee
and also
\be
\label{Mi13}
n_s - 1 \sim - 3\epsilon\, ,\quad \alpha_s \sim - 3 \epsilon^2\, .
\ee
The tensor-to-scalar ratio is given by
\be
\label{S3}
r = 16 \epsilon\, ,
\ee
The Planck analysis \cite{Ade:2013lta, Ade:2013uln} gives
$n_{\mathrm{s}} = 0.9603 \pm 0.0073\, (68\%\,\mathrm{CL})$
and $\alpha_\mathrm{s} = -0.0134 \pm 0.0090\, (68\%\,\mathrm{CL})$
by using the data of the Planck and WMAP \cite{WMAP, Hinshaw:2012aka}.
We should note that the sign of $\alpha_\mathrm{s}$ is negative at
$1.5 \sigma$ level. We also find $r< 0.11\, (95\%\,\mathrm{CL})$.
The result of the BICEP2 experiment, however, gives
$r = 0.20_{-0.05}^{+0.07}\, (68\%\,\mathrm{CL})$ \cite{Ade:2014xna}
(see, e.g., Refs.~\cite{A-A, Mortonson:2014bja} for recent discussions).
If $n_{\mathrm{s}} = 0.9603$, Eq.~(\ref{Mi13}) gives $\epsilon=0.0132$ and
therefore $r=0.2117$, which might be consistent with the observed data.

In case of the late-time $\Lambda$CDM era, the scale factor is
\be
\label{Mi14}
a(t) = A \sinh^{\frac{3}{2}}\left(\alpha t\right)\, ,
\ee
with constants $A$ and $\alpha$, Eq.~ (\ref{Mi8}) has the following form:
\be
\label{Mi15}
0 = \frac{d^2 f}{dt^2} - \frac{3}{2}\alpha \coth \left(\alpha t\right)
\frac{df}{dt} - \frac{3\alpha^2}{\sinh^2 \left( \alpha t\right)} f\, .
\ee
By changing the variables by
\be
\label{Mi16}
x \equiv - \frac{1}{\sinh^2 \left( \alpha t \right)}\, ,
\ee
Eq.~(\ref{Mi15}) becomes hypergeometric differential equation,
\begin{align}
\label{Mi17}
& 0 = x \left( 1 - x \right) \frac{d^2 f}{dx^2} + \left\{ \gamma - \left(
\alpha + \beta + 1
\right) x \right\} \frac{df}{dx} - \alpha \beta f \, , \nn
& \gamma = 7\, ,\quad
\alpha + \beta + 1 = 9\, , \quad
\alpha \beta = - \frac{3}{4} \, ,
\end{align}
whose solutions are given by the hypergeometric functions:
\be
\label{Mi18}
f_1 = \,_2F_1\left(\alpha,\beta,\gamma;x \right)\, ,\quad
f_2 = \left( 1 - x \right)^{1-\gamma} \,_2F_1
\left( 1+ \alpha - \gamma , 1+ \beta - \gamma, 2 - \gamma; x \right)\, .
\ee
We should note that the scalar curvature is given by
\be
\label{Mi19}
R = 9\alpha^2 \left( 3 + \frac{2}{\sinh^2 \left(\alpha t \right)} \right)\, ,
\ee
and therefore
\be
\label{Mi20}
x = \frac{3}{2} - \frac{R}{18\alpha^2}\, .
\ee
Because
\be
\label{Mi21}
\int dx \,_2F_1\left(\alpha,\beta,\gamma;x \right)
= \frac{\gamma -1}{\left( \alpha - 1 \right) \left( \beta - 1 \right)}
\,_2F_1\left(\alpha - 1,\beta - 1,\gamma - 1; x \right) \, ,
\ee
when $\rho=p=C_\phi=0$ as a special case, by using the first solution in
(\ref{Mi18}), we find
\be
\label{Mi22}
F(R) = F_0 \,_2F_1\left(\alpha - 1,\beta - 1,\gamma - 1;
\frac{3}{2}  - \frac{R}{18\alpha^2} \right) \, .
\ee
with a constant $F_0$. The obtained expression (\ref{Mi22}) is not changed from
that in \cite{Nojiri:2009kx} because we have considered the case that
$\rho=p=C_\phi=0$.
When $C_\phi\neq 0$, the form of $F(R)$ is different from that of the standard 
$F(R)$
gravity although the explicit form becomes very complicated.
The explicit form is given in Appendix.
Thus, we demonstrated that realistic dark energy epoch may be obtained from the
mimetic $F(R)$ gravity different from convenient $F(R)$.

We may reconstruct the evolution, which describes both of inflation and
the recent accelerating expansion of the  universe:
\be
\label{Mi22B}
H = H_0 \frac{ 1 + \epsilon \left( \frac{t}{t_0} \right)^2}{1 + \left(
\frac{t}{t_0} \right)^2}\, .
\ee
When $t\to 0$, $H$ behaves as
\be
\label{Mi23}
H \sim H_0 \left( 1 - \left( 1 - \epsilon \right) \left( \frac{t}{t_0}
\right)^2 + \mathcal{O}\left( \left( \frac{t}{t_0} \right)^4 \right) \right)\, 
,
\ee
and $H$ goes to a constant $H_0$ and e-foldings number $N$ is given by
$N\sim H_0 t_0$.
On the other hand, when $t$ is large, we find
\be
\label{Mi24}
H \sim \epsilon H_0 \left( 1 + \left( \frac{1}{\epsilon} - 1 \right) \left(
\frac{t_0}{t} \right)^2
+ \mathcal{O}\left( \left( \frac{t_0}{t} \right)^4 \right) \right)\, ,
\ee
and therefore $H$ goes to a constant $\epsilon H_0$.
It is rather difficult to solve (\ref{Mi8}) explicitly.
One can, however, solve (\ref{Mi8}) when $t\to 0$ or $t\to \infty$.
When $t\to 0$, a solution of (\ref{Mi8}) is given by
\be
\label{Mi25}
f(t) = f_0 \left( 1 - \frac{4\left( 1 - \epsilon \right)}{H_0 t_0} \left(
\frac{t}{t_0} \right) - 2 \left( 1 - \epsilon \right)
\left( \frac{t}{t_0} \right)^2 + \mathcal{O}\left( \left( \frac{t}{t_0}
\right)^3 \right)
\right) \, .
\ee
On the other hand, when $t\to\infty$, we find
\be
\label{Mi26}
f(t) = f_\infty \left( 1 + 2\left( \frac{1}{\epsilon} - 1 \right)
\left( \frac{t_0}{t} \right)^2 + \mathcal{O}\left( \left( \frac{t_0}{t}
\right)^3 \right) \right) \, .
\ee
When $t\to 0$, $R$ is given by
\be
\label{Mi26B}
R = 12 H_0^2 - \frac{12 H_0 \left( 1 - \epsilon\right) t}{t_0^2}
+ \mathcal{O}\left( \left( \frac{t}{t_0} \right)^3 \right) \, ,
\ee
and $f(t)$ in (\ref{Mi25}) has the following form
\be
\label{Mi26C}
f(t) \sim f_0 \left( 1 - \frac{R - 12 H_0^2}{3 H_0^2} \right)\, .
\ee
Then in case that $\rho=p=C_\phi=0$, we find
\be
\label{Mi26D}
F(R) \sim f_0 \left( R - 12 H_0^2 - \frac{\left(R - 12 H_0^2\right)^2}{6 H_0^2}
\right) + \mathrm{const.}\, .
\ee
On the other hand, when $t\to \infty$, we find
\be
\label{Mi26E}
R = 12 \epsilon^2 H_0^2 \left( 1 + 2 \left( \frac{1}{\epsilon} - 1 \right)
\left( \frac{t_0}{t} \right)^2\right) + \mathcal{O}\left( \left( \frac{_0t}{t} 
\right)^4 \right)
\, ,
\ee
and therefore Eq.~(\ref{Mi26}) gives
\be
\label{Mi26F}
f(t) \sim \frac{f_\infty R}{12\epsilon^2 H_0^2} \, ,
\ee
and
\be
\label{Mi26G}
F(R) \sim \frac{f_\infty R^2}{24\epsilon^2 H_0^2} + \mathrm{const.} \, .
\ee
The obtained expressions in (\ref{Mi26D}) and (\ref{Mi26G}) are not changed
from those in the standard $F(R)$ gravity because we are considering the case 
that
$\rho=p=C_\phi=0$.
When $C_\phi\neq 0$, the form of $F(R)$ is different from that if the standard 
$F(R)$
gravity. We do not present it explicitly because it has very complicated form.
Thus, the possibility to unify inflation with dark energy in the mimetic
$F(R)$ gravity is possible following the first proposal of Ref.~\cite{uni}.

We finally consider the bouncing universe (for review, see \cite{rb}),
whose scale factor is given by
\be
\label{Mi27}
a(t) = \e^{\alpha t^2}\, ,
\ee
with a constant $\alpha$.
Then Eq.~(\ref{Mi8}) has the following form:
\be
\label{Mi28}
0 = \frac{d^2 f(t)}{dt^2} - 2\alpha t \frac{d f(t)}{dt} + 4 \alpha f(t)\, ,
\ee
whose solutions are given by
\be
\label{Mi29}
f_1(t) = f_1^{(0)}\left( t^2 - \frac{1}{2\alpha} \right)\, ,\quad
f_2(t) = f_2^{(0)}\left( t^2 - \frac{1}{2\alpha} \right) \int^t dt' \frac{\e^{
- \alpha {t'}^2}}{
\left( 2 \alpha {t'}^2 - 1 \right)}\, .
\ee
Here $f_1^{(0)}$ and $f_2^{(0)}$ are constants.
Because it is difficult to give the explicit expression for (\ref{Mi9}), we
again consider the simple case where $\rho=p=C_\phi=0$.
Then we find
\be
\label{Mi30}
F(R) = F_0 \left( \frac{R^2}{\alpha} - 72 R + 144 \alpha \right)\, ,
\ee
with a constant $F_0$, which reproduces the expression in \cite{Bamba:2013fha}.
Thus, the possible occurrence of bouncing universe is also demonstrated.

\section{Discussion.}

In summary, following the idea of Ref.~\cite{Chamseddine:2013kea} we
presented mimetic $F(R)$ gravity. It is demonstrated that such theory
being  conformally invariant one, admits the inflation, dark
energy, unification of inflation with dark energy as well as bounce much
in the same way as convenient $F(R)$ gravity. Note that interpretation of
mimetic dark matter is the same as in usual mimetic gravity. Note also that 
convenient $F(R)$ gravity is known to be ghost-free. Usually, the addition of 
Lagrange multiplier constraint does not violate the ghost-free property. Hence, 
we expect that mimetic $F(R)$ gravity under consideration is ghost-free. 
Nevertheless, this conjecture should be verified in hamiltonian formulation.

It might be interesting if we add a potential term for $\phi$ to the action
(\ref{EHaction})
\be
\label{EHphiaction}
S=\int d^4 x \sqrt{-g\left({\hat g}_{\mu\nu}, \phi \right)}
\left( \frac{R\left(g\left({\hat g}_{\mu\nu}, \phi
\right)_{\mu\nu}\right)}{2\kappa^2} - V(\phi) + 
\mathcal{L}_\mathrm{matter}\right)\, .
\ee
Due to the constraint in (\ref{Mi2B}),one can identify $\phi$ with time in
the FRW space-time (\ref{FRW}).
Therefore, with the action (\ref{EHphiaction}) one effectively obtains
time-dependent energy density, which could give quite an interesting cosmology.

In \cite{Chamseddine:2014vna}, it has been proposed that instead of
parameterizing the metric as in (\ref{Mi1}), we may impose the condition 
(\ref{Mi2B}) via
the Lagrange multiplier field $\lambda$ addition:
\be
\label{EHphiactionlambda}
S=\int d^4 x \sqrt{-g} \left( \frac{R\left(g_{\mu\nu}\right)}{2\kappa^2}
  - V(\phi) + \lambda \left( g^{\mu\nu} \partial_\mu \phi \partial_\nu \phi + 1 
\right)
+ \mathcal{L}_\mathrm{matter}\right)\, .
\ee
Anyway, one can identify $\phi$ with time in the FRW space-time (\ref{FRW}),
and the potential $V(\phi)=V(t)$ gives a time-dependent energy density.
Hence, one  may realize arbitrary evolution of the expansion of the universe by 
adjusting
$V(\phi)$.

Then instead of (\ref{EHphiactionlambda}), we may consider the following action
of mimetic $F(R)$ gravity with scalar potential:
\be
\label{FRphilambda}
S=\int d^4 x \sqrt{-g} \left( F \left(R\left(g_{\mu\nu}\right)\right)
  - V(\phi) + \lambda \left( g^{\mu\nu} \partial_\mu \phi \partial_\nu \phi + 1
\right) + \mathcal{L}_\mathrm{matter}\right)\, .
\ee
This action is of the sort of modified gravity with Lagrange multiplier
constraint \cite{mat}.
By the variation of the action (\ref{FRphilambda}) with respect to the
metric, one obtains
\begin{align}
\label{Mi31}
0=& \frac{1}{2} g_{\mu\nu} F\left(R\right) - R_{\mu\nu} F'\left(R\right)
+ \nabla_\mu \nabla_\nu F'\left(R\right) - g_{\mu\nu} \Box F'\left(R\right) \nn
&+ \frac{1}{2} g_{\mu\nu} \left\{  - V(\phi)
+ \lambda \left( g^{\rho\sigma} \partial_\rho \phi \partial_\sigma
\phi + 1 \right)
\right\} - \lambda \partial_\mu \phi \partial_\nu \phi + \frac{1}{2} 
T_{\mu\nu}\, .
\end{align}
On the other hand, by the variation with respect to $\phi$, we get
\be
\label{Mi32}
0= -2 \nabla^\mu \left( \lambda \partial_\mu \phi \right) - V'(\phi)\, .
\ee
By construction, the variation with respect to $\lambda$ gives,
\be
\label{Mi33}
0 = g^{\rho\sigma} \partial_\rho \phi \partial_\sigma \phi + 1 \, .
\ee
In the FRW space-time, assuming that $\phi$ depends only on time
coordinate $t$,
Eqs.~(\ref{Mi31}), (\ref{Mi32}), and (\ref{Mi33}) are:
\begin{align}
\label{Mi34}
0 =& - F(R) + 6 \left( \dot H + H^2 \right) F'(R) - 6H \frac{d F'(R)}{dt}
  - \lambda\left( {\dot \phi}^2 + 1 \right) + V(\phi) + \rho\, , \\
\label{Mi35}
0=& F(R) - 2 \left(\dot H + 3H^2 \right) + 2 \frac{d^2 F'(R)}{dt^2}
+ 4H \frac{d F'(R)}{dt} - \lambda\left( {\dot \phi}^2 - 1 \right) - V(\phi) + p 
\, , \\
\label{Mi36}
0 =& 2 \frac{d}{dt} \left( \lambda \dot \phi \right) + 6 H \lambda \dot \phi
  - V'(\phi)\, , \\
\label{Mi37}
0 =& {\dot\phi}^2 - 1\, .
\end{align}
Eq.~(\ref{Mi36}) shows that  $\phi$ can be identified as the time
coordinate:
$\phi=t$.
Then Eq.~(\ref{Mi35}) can be rewritten as
\be
\label{Mi38}
0 = F(R) - 2 \left(\dot H + 3H^2 \right) + 2 \frac{d^2 F'(R)}{dt^2}
+ 4H \frac{d F'(R)}{dt} - V(\phi=t) + p \,  .
\ee
Therefore in case that matter contribution can be neglected ($p=\rho=0$),
we find
\be
\label{Mi39}
V(t) = F(R) - 2 \left(\dot H + 3H^2 \right) + 2 \frac{d^2 F'(R)}{dt^2}
+ 4H \frac{d F'(R)}{dt} \, .
\ee
Taking $\rho=0$,  Eq.~(\ref{Mi34}) can be solved with respect to
$\lambda$:
\be
\label{Mi40}
\lambda (t) = - \frac{1}{2}F(R)
+ 3 \left( \dot H + H^2 \right) F'(R) - 3H \frac{d F'(R)}{dt}
\ee
Hence, Eq.~(\ref{Mi36}) is automatically satisfied.
Then by choosing the potential $V(\phi)$, we can construct a model which
reproduces arbitrarily given evolution $H$.
Doing this one can work with viable $F(R)$ gravities \cite{review}.
For instance, we can easily get the same accelerating behavior as in the 
previous section.
Note that for given scalar potential one can reconstruct  function $F(R)$ to 
reproduce the
arbitrarily given universe evolution. Of course, also using the arbitrary form 
of the function $V(\phi)$ one can get accelerating universe for specific $F(R)$ 
gravity which basically does not admit such evolution or admits totally 
different accelerating expansion.
The detailed study of accelerating cosmology in mimetic $F(R)$ gravity with 
scalar potential will be done elsewhere.

Final remark is in order. The method to make the conformal symmetry to be
the internal property of theory via the corresponding parametrization of the
metric seems to be quite wide. It would be interesting to apply it to
higher-derivative (quantum) dilaton gravity of Ref.~\cite{eli} what may
make the bridge between quantum gravity and mimetic dark matter.

\section*{Acknowledgements.}

The work by S.N. is supported by the JSPS Grant-in-Aid for Scientific
Research (S) \# 22224003 and (C) \# 23540296.
The work by S.D.O. is supported in part by MINECO (Spain), project
FIS2010-15640, and by the CPAN Consolider Ingenio Project.

\appendix

\section{Explicit form of $F(R)$ when $C_\phi \neq 0$}

In this appendix, we give an explicit form of $F(R)$ when $C_\phi \neq 0$ for 
the model in
(\ref{Mi14}).
First we rewrite Eq.~(\ref{Mi9}) in terms of $x$ in (\ref{Mi16}) instead of $t$ 
as follows,
\begin{align}
\label{AMi9}
F'\left(R \left(t\right)\right) =& - f_1 (x) \int^x dt' \frac{ \gamma(x')
f_2(x') }{\tilde W\left(f_1(s'),f_2(x')\right)}
+ f_2 (x) \int^x dx' \frac{ \gamma(x') f_1(x') }{
\tilde W\left(f_1(x'),f_2(x')\right)} \, , \nn
\tilde W\left(f_1,f_2\right) \equiv & f_1(x) f_2'(x) - f_2(x) f_1'(x)\, .
\end{align}
The scale factor $a$ is also expressed in terms of $x$,
\be
\label{Ap1}
a = A \left( - \frac{1}{x} \right)^{\frac{3}{4}}\, .
\ee
For simplicity, we consider the case $\rho=p=0$. Then by using (\ref{Mi18}) and 
formula
\be
\label{Ap2}
\frac{d\,_2F_1\left(\alpha,\beta,\gamma;x \right)}{dx}
= \frac{\alpha \beta}{\gamma}
\,_2F_1\left(\alpha+1,\beta+1,\gamma+1;x \right)\, ,
\ee
we find more explicit form of $F(R)$ corresponding to (\ref{Mi14}) as follows,
\begin{align}
\label{Ap3}
F(R) =& \frac{C_\phi}{A^3} \int^{\frac{3}{2} - \frac{R}{18\alpha^2}} dx' \left( 
- x'
\right)^{\frac{9}{4}} \left[ - \,_2F_1\left(\alpha,\beta,\gamma; \frac{3}{2} - 
\frac{R}{18\alpha^2}
\right) \left( 1 - x' \right)^{1-\gamma} \,_2F_1
\left( 1+ \alpha - \gamma , 1+ \beta - \gamma, 2 - \gamma; x' \right) \right. 
\nn
& \left. + \left( 1 - \frac{3}{2} + \frac{R}{18\alpha^2} \right)^{1-\gamma} 
\,_2F_1
\left( 1+ \alpha - \gamma , 1+ \beta - \gamma, 2 - \gamma; \frac{3}{2} - 
\frac{R}{18\alpha^2}
\right)\,_2F_1\left(\alpha,\beta,\gamma; x' \right) \right] \nn
& \times \left[ \,_2F_1\left(\alpha,\beta,\gamma; x' \right)
\left( \left( 1 - \gamma \right) \left( 1 - x' \right)^{-\gamma} \,_2F_1
\left( 1+ \alpha - \gamma , 1+ \beta - \gamma, 2 - \gamma; x' \right) \right. 
\right. \nn
& \left.
+ \frac{\left( 1+ \alpha - \gamma \right) \left( 1 + \beta - \gamma \right)}{2 
- \gamma}
\left( 1 - x' \right)^{1-\gamma} \,_2F_1
\left( 2+ \alpha - \gamma , 2+ \beta - \gamma, 3 - \gamma; x' \right) \right) 
\nn
& \left. - \frac{\alpha \beta}{\gamma} \,_2F_1\left(\alpha+1,\beta+1,\gamma+1 ; 
x' \right)
\left( 1 - x' \right)^{1-\gamma} \,_2F_1
\left( 1+ \alpha - \gamma , 1+ \beta - \gamma, 2 - \gamma; x' \right) 
\right]^{-1}\, .
\end{align}
Here we have used (\ref{Mi20}).
Then the obtained mimetic $F(R)$ (\ref{Ap3}) is completely different from that 
in case $C_\phi =0$ in
(\ref{Mi22}).

\end{document}